\documentclass[12pt]{article}
\usepackage[T1]{fontenc}
\usepackage{xcolor}
\usepackage{amsmath,amsfonts,amsthm,amssymb}
\usepackage{mathtools}
\usepackage{bm}
\usepackage{bbm} 
\usepackage{algorithm}
\usepackage{algpseudocode}
\usepackage{longtable}
\usepackage{caption}
\usepackage{subcaption}

\newcommand{\Cat}{\operatorname{Cat}}
\newcommand{\EMA}{\operatorname{EMA}}
\newcommand{\MACD}{\operatorname{MACD}}

\newcommand{\RSI}{\operatorname{RSI}}
\newcommand{\ASPI}{\operatorname{ASPI}}

\title{Portfolio optimisation via the Heston model calibrated to real asset data}

\providecommand{\keywords}[1]
{
  \small	
  \textbf{Keywords: } #1
}

\author{Jarosław Gruszka$^{1,2}$, Janusz Szwabiński$^1$}
\date{{\small
    $^1$ Hugo Steinhaus Center at The Faculty of Pure and Applied Mathematics\\
    Wrocław University of Science and Technology\\
    Wyspiańskiego 27, 50-370 Wrocław, Poland\\}
    \vspace{0.3cm}{\small
    $^2$ corresponding author: \textit{jaroslaw.gruszka@pwr.edu.pl}
    }
}

\begin{document}

\maketitle

\begin{abstract}
    The debate between active and passive investment strategies has been ongoing for many years and is far from being over. In this paper, we show that the choice of an optimal portfolio management strategy depends on an investment climate, which we measure via the parameters of the Heston model calibrated to the real stock market data. Depending on the values of those parameters, the passive strategy may namely outperform the active ones or vice versa. The method is tested on three stock market indices: S\&P500, DAX and WIG20.
\end{abstract}


\section{Introduction}

Present-day financial markets provide a multitude of ways of investing money --- from classical stocks to more complex derivative instruments. Selecting financial instruments that are going to fit one's investment goal and risk tolerance is a critical part of building financial portfolio. There is however also another factor, equally if not more important and that is selecting the investment strategy. By the investment strategy we understand investor's approach to changes in the composition of their portfolio --- in particular by selling some of the assets and buying other ones, hoping for a greater return on the entire investment. There is a very wide range of factors that can influence once's investment decisions --- most investors take into consideration the intrinsic characteristics of a given asset, like the financial performance of the company of which shares they want to buy \cite{izuddin_impact_2021}, but it has also been proven that investor's emotions play a significant role in the decision making process \cite{subrahmanyam_behavioural_2008}. Arguably, however, one of the most important factors that investor's take into account while making their decisions is the history of the price of a given asset and its fluctuations. The past market value of any kind of financial resource and the history of how this value was changing over time provides a good frame of reference and can be treated as an indication of what sort of values this very resource might attain in the future. It needs to be kept in mind however, that the prices of financial assets nowadays are nothing short of ever-changing and seemingly random. This is why whenever an analytical approach is taken for modelling prices of financial resources, almost always the model using a probabilistic approach is selected, like the Geometric Brownian Motion \cite{hull_options_2018} (often called GBM in short) for example. A big part of modern research in the field of financial mathematics and econophysics is knowing deep details about those models as well as the actual behaviour of real assets in various circumstances. The aforementioned GBM process has been recently heavily studied in respect to its property called ergodicity \cite{stojkoski_geometric_2021, stojkoski_ergodicity_2022, vinod_time-averaging_2022}, while the recent real-data studies unravel many surprising results about linear and non-linear market correlations \cite{haluszczynski_linear_2017} and the role of market volatility \cite{valenti_stabilizing_2018}.

The apparent randomness and unpredictability of the prices of assets is one of the reasons why most investors tend to re-consider the content of their portfolio throughout the time of the investment. This is what we call "active" portfolio management. Changing prices of the financial assets are often considered an opportunity to increase the value of the portfolio and the simplest way to do it is by selling the assets which, based on the price history, seem to be overvalued (expensive) and use the money to buy new assets, which are undervalued (cheap) at that moment. Ideally, investors seek the moment of selling their assets just before their prices start to fall and buying them before prices start to  increase. It is obviously not easy to assess when those moments take place, so various indicators, like MACD \cite{appel_technical_2005} or RSI \cite{wilder_new_1978}, have been designed to help find suitable moments for making transactions. 

Investing based on trading indicators is always related to the risk of performing transactions on false positive signals and consequently --- incurring a loss \cite{abbey_is_2012}. Hence, a strategy proposing not to do any changes in the portfolio throughout its planned lifetime is preferred by some investors. This is called the "buy-and-hold" strategy or a "passive" investment portfolio \cite{gruszka_best_2020}. Those who follow passive investing argue that for every investor who successfully run their portfolio in such a way that they significantly outperformed other members of the market there is one (or more) which fell behind others, and the list of "successful" investors changes constantly, so statistically speaking it is a question of sheer luck if one is able to "beat the market" (i.e. perform such reallocation in the content of the portfolio that its growth is faster than the general, average growth of all assets traded on the market, measured e.g. by a stock index).

In one of our previous papers \cite{gruszka_advanced_2021} we have shown that the performance of different portfolio management strategies within the Heston market model depends on the values of its parameters. In this work, we are going to exploit that result in order to find an optimal investment strategy for a given set of assets. The idea is quite simple: we can check the  ``character'' of the assets by calibrating the Heston model to their trajectories. Based on the estimated values of the parameters we can pick the appropriate strategy to manage the portfolio.

The paper is structured as follows. In section \ref{sec:models_and_methods} we briefly introduce the Heston model without and with jumps, present the estimation procedure of its parameters and finally summarize the investment strategies of interest. In Sec.~\ref{sec:real_data_results}, the method will be tested on three stock indices: S\&P500, DAX and WIG20. Conclusions will be drawn in the last section.

\section{Models and methods}
\label{sec:models_and_methods}

The Bayesian-inference-based estimation framework \cite{lindley_bayes_1972, ohagan_kendalls_1994} will be used for the purpose of this paper. In this section, we briefly introduce the Heston model and sketch the whole estimation procedure, we also present the definitions of our investment strategies. The interested reader may refer to Ref.~\cite{gruszka_parameter_2022} for more details on the estimation method.   

\subsection{Heston model}

The basic Heston model~\cite{heston_closed-form_1993} assumes that $S(t)$, the price of the asset, is determined by a stochastic process,
\begin{equation}
dS(t) = \mu S(t) dt + \sqrt{v(t)} S(t) d B^S(t),
\label{eq:Heston_stock}
\end{equation}
where the volatility $v(t)$ is given by a Feller square-root or CIR process~\cite{cox_theory_1985},
\begin{equation}
	dv(t) = \kappa(\theta - v(t))dt + \sigma \sqrt{v(t)} S(t) d B^v(t).
	\label{eq:Heston_vol}
\end{equation}
The parameter $\mu$ represents the drift of the stock price.  $\theta$ is the long-term average from which the volatility diverges and to which it then comes back (due to the mean-reversion feature of the CIR process) and $\kappa$ stands for the rate of those fluctuations (the bigger $\kappa$, the longer it takes to come back to $\theta$).  Parameter $\sigma$ is called the volatility-of-the-volatility and  is generally responsible for the "scale" of the randomness of the volatility process. 
The Wiener processes $B^s(t)$ and $B^v(t)$ in the above equations may be correlated with an instantaneous correlation coefficient $\rho$,
\begin{equation}
	\label{eq:Heston_rho}
	dB^S(t) dB^v(t) = \rho dt.
\end{equation}
To complete the set-up, deterministic initial conditions for $S$ and $v$ need to be specified: 
\begin{gather}
	\label{eq:Heston_ic}
	S(0) = S_0 > 0,\\
	v(0) = v_0 >0.
\end{gather}
Here, $S_0$ represents the initial price of an asset at time $t=0$, and $v_0$  is the value of the market volatility at that point in time.

\subsubsection{Merton-style jumps}

Asset price trajectories generated by the Heston model are characterised by continuity, which is generally inconsistent with the actual stock evolution, since jumps occur  often in the real asset paths. The behaviour of the model may be improved by adding Merton log-normal jumps~\cite{merton_option_1976}. This is formally done by adjusting Equation \eqref{eq:Heston_stock} to become
\begin{equation}
    \label{eq:Heston_stock_jump}
    dS(t) = \mu S(t) dt + \sqrt{v(t)} S(t) d B^S(t) + (e^{Z(t)}-1) S(t) dq(t),
\end{equation}
where $Z(t)$ is a series of i.i.d. normally distributed random variables with mean $\mu^J$ and standard deviation $\sigma^J$, whereas $q(t)$ is a Poisson counting process with constant intensity $\lambda$. The added term turns the Heston model into the Bates model \cite{bates_jumps_1996}. The above extension has an easy real-life interpretation. Namely, $e^{Z(t)}$ is the actual (absolute) rate of the difference between the price before the jump at time $t$ and right after it, i.e. $S(t-)\cdot e^{Z(t)} = S(t+)$. If for example, for a given $t$, $e^{Z(t)} \approx 0.85$, that means the stock experienced $\sim 15\%$ drop in value at that moment.

\subsubsection{Simulation procedure}

Given some arbitrary set of equidistant points in time, $\mathcal{T}=\{t_k\}_{k=0}^n = \{k\Delta t\}_{k=0}^n$, where $\Delta t = t_k - t_{k-1}$, the random paths of the pair $(S(t),v(t))$ will be generated with the  Euler-Maruyama discretisation scheme~\cite{kloeden_numerical_1992}.  The stock price equation \eqref{eq:Heston_stock} can be discretised as 
\begin{multline}
	\label{eq:Heston_s_disc}
	S(k\Delta t) - S\Big((k-1)\Delta t\Big) = \mu S\Big((k-1)\Delta t\Big)\Delta t + \\
	S\Big((k-1)\Delta t\Big) \sqrt{v\Big((k-1)\Delta t\Big)} \varepsilon ^S(k\Delta t) \sqrt{\Delta t},
\end{multline}
where $k \in \{1, \cdots, n\}$ and $\varepsilon^S$ is a series of $n$ i.i.d. standard normal random variables. To highlight the ratio between two consecutive values of the stock price, Eq.~\eqref{eq:Heston_s_disc} is often re-written as
\begin{equation}
	\label{eq:Heston_s_disc_ret}
	R(k\Delta t) \equiv \frac{S(k\Delta t)}{S\Big((k-1)\Delta t\Big)}  = \mu \Delta t +1  + \sqrt{v\Big((k-1)\Delta t\Big)} \varepsilon ^S(k\Delta t) \sqrt{\Delta t}.
\end{equation}
The same discretisation scheme can be applied to Eq.~\eqref{eq:Heston_vol}:

\begin{multline}
	\label{eq:Heston_v_disc}
	v(k\Delta t) - v\Big((k-1)\Delta t\Big) = \kappa\Bigg(\theta - v\Big((k-1)\Delta t\Big)\Bigg)\Delta t + \\
	\sigma \sqrt{v\Big((k-1)\Delta t\Big)} \varepsilon ^v (k\Delta t) \sqrt{\Delta t}.
\end{multline}
If $\rho = 0$, then $\varepsilon^v$ in the above formula  is also a series of $n$ i.i.d. standard normal random variables. However, if $\rho \neq 0$, then --- to ensure the proper dependency between $S$ and $v$ --- we take 

\begin{equation}
	\label{eq:eps_v}
	\varepsilon ^v (k\Delta t) = \rho \varepsilon ^S(k\Delta t) + \sqrt{1-\rho^2} \varepsilon^{add} (k\Delta t)
\end{equation}
where $\varepsilon^{add}$ is an additional series of $n$ i.i.d. standard normal random variables, which are "mixed" with the ones from $\varepsilon^S$ and hence --- become dependent on them.

\subsection{Estimation framework}

The estimation framework for the parameters of the Heston model from the process of the basic instrument only (!) which we present here has been created according to the guidelines from Ref.~\cite{johannes_chapter_2010}. Our approach is a mixture of Bayesian regression~\cite{ohagan_kendalls_1994} and the particle filtering method~\cite{johannes_optimal_2009, christoffersen_volatility_2007}. While the first approach will be utilised to estimate the parameters of the regular model, the latter will be used to assess the volatility and the jump-related parameters. In this section, the highlights of the framework will be presented. Please refer Ref.~\cite{gruszka_parameter_2022} for further details.

\subsubsection{Estimation of $\mu$}

In general, the Bayesian regression is a type of conditional modelling in which the distribution of one variable is described by a combination of other variables, with the goal of obtaining the posterior probability of the regression coefficients and allowing the out-of-sample prediction of the regressand conditional on observed values of the regressors.

In order to estimate the drift $\mu$, we  first transform the ratio of two consecutive values of the asset price to the following linear regression form:
\begin{equation}
	\label{eq:Heston_s_est2}
	y^S(k\Delta t) = \eta x^S(k\Delta t)  + \varepsilon ^S(k\Delta t). 
\end{equation}
where
\begin{eqnarray}
	\label{eq:y_s}
	y^S(k\Delta t) & = & \frac{1}{\sqrt{v\Big((k-1)\Delta t\Big)}\sqrt{\Delta t}}R(k\Delta t),\\
	x^S(k\Delta t) & = & \frac{1}{\sqrt{v\Big((k-1)\Delta t\Big)}\sqrt{\Delta t}},\nonumber
\end{eqnarray}
and $\eta =\mu\Delta t + 1$ is a helper variable related to the drift.

To apply the Bayesian regression framework,  we collect all discrete values of $y^S(t)$ and $x^S(t)$ into $n$-element column vectors:
\begin{equation}
	\label{eq:y_s_vec}
	\mathbf{y}^S = \frac{1}{\sqrt{\Delta t}}
	\begin{bmatrix}
		\frac{R(\Delta t)}{\sqrt{v(0)}} &
		\frac{R(2 \Delta t)}{\sqrt{v(\Delta t)}} & 
		\ldots &
		\frac{R(n\Delta t)}{\sqrt{v\big((n-1)\Delta t\big)}}
	\end{bmatrix}^\prime ,
\end{equation}
and 
\begin{equation}
	\label{eq:x_s_vec}
	\mathbf{x}^S = \frac{1}{\sqrt{\Delta t}}
	\begin{bmatrix}
		 \frac{1}{\sqrt{v(0)}} &
		 \frac{1}{\sqrt{v(\Delta t)}} & 
		 \ldots &
		 \frac{1}{\sqrt{v\big((n-1)\Delta t\big)}}
	\end{bmatrix} ^\prime ,
\end{equation}
where the prime symbol is used for the transpose.

Assuming a prior distribution for $\eta$ to be normal with mean $\mu_0^\eta$ and standard deviation $\sigma_0^\eta$, according to the Bayesian regression the posterior distribution for $\eta$ will also be normal. Thus, sample realisations of $\eta$ can be drawn from
\begin{equation}
	\label{eq:eta_i_bayes}
	\eta_i \sim \mathcal{N}\left(\mu^\eta, \frac{1}{\sqrt{\tau^\eta}}\right)
\end{equation}
where $i$ indicates the $i$-th sample from the posterior distribution of $\eta$, $\tau^\eta$ is the precision (i.e. the inverse of variance) of the posterior distribution,
\begin{equation}
\label{eq:tau_eta}
\tau^\eta = \left(\bm{x}^S\right)' \cdot \bm{x}^S + \tau_0^\eta,
\end{equation}
and the mean $\mu^\eta$ of the posterior distribution is given by
\begin{equation}
\label{eq:mu_eta}
\mu^\eta = \frac{1}{\tau^\eta} \left( \tau_0^\eta \mu_0^\eta + \left(\bm{x}^S\right)' \cdot \bm{x}^S \hat{\eta}\right).
\end{equation}
In the above equations, $\tau_0^\eta= \frac{1}{\left(\sigma_0^\eta\right)^2}$ is the precision of the prior distribution and $\hat{\eta}$ stands for the classical ordinary-least-squares (OLS) estimator of $\eta$,
\begin{equation}
\label{eq:eta_hat}
\hat{\eta} = \left(\left(\bm{x}^S\right)' \cdot \bm{x}^S\right)^{-1}\left(\bm{x}^S\right)'\bm{y}^S.
\end{equation}

Having a realization of $\eta$,  we can quickly turn it into a realisation of the drift parameter $\mu$:
\begin{equation}
	\label{eq:mu_i}
	\mu_i = \frac{\eta_i-1}{\Delta t}.
\end{equation}

\subsubsection{Estimation of $\kappa$, $\theta$ and $\sigma$}

A similar exercise is required to estimate the parameters related to the volatility process~\eqref{eq:Heston_vol}, i.e. $\kappa$, $\theta$ and $\sigma$.
Introducing the new variable
\begin{equation}
	\label{eq:beta_vec}
	\bm{\beta} = 
	\begin{bmatrix}
		 \beta_1 \\ 
		 \beta_2
	\end{bmatrix}
        = 
	\begin{bmatrix}
		 \kappa\theta\Delta t \\ 
		 1-\kappa\Delta t
	\end{bmatrix}
\end{equation}
and the vectors
\begin{equation}
	\label{eq:y_v_vec}
	\mathbf{y}^v = \frac{1}{\sqrt{\Delta t}}
	\begin{bmatrix}
		 \frac{v(2\Delta t)}{\sqrt{v(\Delta t)}} &
		 \frac{v(3\Delta t)}{\sqrt{v(2\Delta t)}} & 
		 \ldots &
		 \frac{v(n\Delta t)}{\sqrt{v\big((n-1)\Delta t\big)}}
	\end{bmatrix}^\prime,
\end{equation}
\begin{equation}
	\label{eq:x_1_v_vec}
	\mathbf{x}_1^v = \frac{1}{\sqrt{\Delta t}}
	\begin{bmatrix}
		 \frac{1}{\sqrt{v(\Delta t)}} &
		 \frac{1}{\sqrt{v(2\Delta t)}} & 
		 \ldots & 
		 \frac{1}{\sqrt{v\big((n-1)\Delta t\big)}}
	\end{bmatrix}^\prime,
\end{equation}
\begin{equation}
	\label{eq:x_2_v_vec}
	\mathbf{x}_2^v = 
	\frac{1}{\sqrt{\Delta t}}
	\begin{bmatrix}
		 \sqrt{v(\Delta t)} &
		 \sqrt{v(2\Delta t)} & 
		 \ldots &
		 \sqrt{v\big((n-1)\Delta t\big)}
	\end{bmatrix}^\prime,
\end{equation}
will  allow us to write down the original volatility equation~\eqref{eq:Heston_vol} in form of a linear regression (see Ref.~\cite{gruszka_parameter_2022} for more details),
\begin{equation}
	\label{eq:v_reg}
	\mathbf{y}^v = \mathbf{X}^v\bm{\beta} + \sigma\bm{\varepsilon}^v
\end{equation}
where
\begin{equation}
	\mathbf{X}^v = 
	\begin{bmatrix}
		 \mathbf{x}_1^v & \mathbf{x}_2^v
	\end{bmatrix},~~~
 \bm{\varepsilon}^v = 
	\begin{bmatrix}
		 \varepsilon^v(\Delta t) &
		 \varepsilon^v(2\Delta t) &
		 \ldots &
		 \varepsilon^v\big((n-1)\Delta t\big)
	\end{bmatrix}.
\end{equation}

Again, assuming a multivariate normal prior with the mean $\bm{\mu}_0$ and the precision $\bm{\Lambda}_0$, the conjugate posterior distribution will be also a multivariate normal,
\begin{equation}
	\label{eq:beta_i_bayes}
	\bm{\beta}_i \sim \mathcal{N}(\bm{\mu}^\beta, \sigma_{i-1}^2(\bm{\Lambda}^\beta)^{-1}),
\end{equation}
with the mean and precision given by
\begin{equation}
	\label{eq:mu_beta}
	\bm{\mu}^\beta = \left(\bm{\Lambda}^\beta\right)^{-1} \left( \bm{\Lambda}_0^\beta \bm{\mu}_0^\beta + \left(\bm{X}^v\right)' \cdot \bm{X}^v \hat{\bm{\beta}}\right)
\end{equation}
and
\begin{equation}
	\label{eq:Lambda_beta}
	\bm{\Lambda}^\beta = \left(\bm{X}^v\right)' \cdot \bm{X}^v + \bm{\Lambda}_0^\beta,
\end{equation}
respectively. $\hat{\bm{\beta}}$ is the standard OLS estimator of $\bm{\beta}$,
\begin{equation}
\label{eq:beta_hat}
\hat{\bm{\beta}} = \left(\left(\bm{X}^v\right)' \cdot \bm{X}^v\right)^{-1}\left(\bm{X}^v\right)'\bm{y}^v.
\end{equation}
Having the realisations of $\bm{\beta}_i $, we get from Eq.~\eqref{eq:beta_vec}
\begin{eqnarray}
	\kappa_i & = &  \frac{1 - \bm{\beta}_{i,2}}{\Delta t},\\
	\theta_i & = & \frac{\bm{\beta}_{i,1}}{\kappa_i\Delta t}.\nonumber
\end{eqnarray}

Note that the realization of $\sigma$ appears in Eq.~\eqref{eq:beta_i_bayes}, but it has not been defined yet. Since the distribution of  $\bm{\beta}$ depends on $\sigma$ and vice versa, we suggest taking the realisation of $\sigma$ from the previous iteration step (indicated by the $i-1$ subscript). 
The most common approach to the estimation of $\sigma$ is assuming the inverse-gamma prior distribution for $\sigma^2$. If the parameters of the prior distribution are $a_0^\sigma$ and $b_0^\sigma$, then the conjugate posterior distribution is also inverse gamma 
\begin{equation}
\label{eq:sigma_i_bayes}
(\sigma_i)^2 \sim \mathcal{I}\mathcal{G}\left(a^\sigma, b^\sigma\right),
\end{equation}
where 
\begin{equation}
	\label{eq:a_sigma}
	a^\sigma = a_0^\sigma + \frac{n}{2}
\end{equation}
and

\begin{equation}
	\label{eq:b_sigma}
	b^\sigma = b_0^\sigma + \frac{1}{2}\left(\left(\bm{y}^v\right)' \cdot \bm{y}^v 
											 + \left(\mu_0^\beta\right)' \Lambda_0^\beta \mu_0^\beta 
											 - \left(\mu^\beta\right)' \Lambda^\beta \mu^\beta \right).
\end{equation}

\subsubsection{Estimation of $\rho$}

For the estimation of $\rho$ we follow an approach presented in \cite{jacquier_bayesian_2004}. Again, after introducing the new variables,
\begin{equation}
\label{eq:psi_omega}
    \psi = \sigma\rho,~~~ \omega = \sigma^2(1-\rho^2),
\end{equation}
we will be able to write down the equations~\eqref{eq:Heston_stock}-\eqref{eq:Heston_rho} defining the model in form of a linear regression:
\begin{equation}
    \label{eq:rho_reg2}
    e_2^\rho(k\Delta t) = \psi e_1^\rho (k\Delta t) + \sqrt{\omega} \varepsilon^{add} (k\Delta t),
\end{equation}
where $e_2^\rho$ and $e_1^\rho$ are the residuals for the volatility and stock price, respectively (see Eqs.~(46) and~(45) in Ref.~\cite{gruszka_parameter_2022} for their definition).

We will assume the prior distributions to be normal in case of $\psi$ and inverse gamma for $\omega$. From the Bayesian regression model it follows, that the realizations of $\omega_i$ and $\psi_i$ will be sampled from the following posteriors:
\begin{eqnarray}
	\omega_i & \sim & \mathcal{I}\mathcal{G}\left(a^\omega, b^\omega\right),\\
 	\psi_i   & \sim & \mathcal{N}\left(\mu^\psi, \frac{\sqrt{\omega_i}}{\sqrt{\tau^\psi}}\right),\nonumber
\end{eqnarray}
with
\begin{eqnarray}
	a^\omega & = & a_0^\omega + \frac{n}{2},\\
	b^\omega & = & b_0^\omega + \frac{1}{2}\left(\mathbf{A}_{22}^\rho - \frac{(\mathbf{A}_{12}^\rho)^2}{\mathbf{A}_{11}^\rho}\right),\\
	\mu^\psi & = & \frac{\mathbf{A}_{12}^\rho + \mu_0^\psi\tau_0^\psi}{\mathbf{A}_{11}^\rho + \tau_0^\psi},\\
 	\tau^\psi & = & \mathbf{A}_{11}^\rho + \tau_0^\psi
\end{eqnarray}
and
\begin{eqnarray}
  	\mathbf{A}^\rho & = & (\mathbf{e}^\rho)' \cdot \mathbf{e}^\rho,\\
  	\mathbf{e}^\rho & = & \left[\mathbf{e}_1^\rho ~ \mathbf{e}_2^\rho\right],\\
	\mathbf{e}_1^\rho &  = &
	\begin{bmatrix}
		 e_1^\rho(\Delta t) & 
		 e_1^\rho(2\Delta t) & 
		 \ldots & 
		 e_1^\rho\left(n\Delta t\right)
	\end{bmatrix}^\prime,\\
	\mathbf{e}_2^\rho & = &
	\begin{bmatrix}
		 e_2^\rho(\Delta t) & 
		 e_2^\rho(2\Delta t) & 
		 \ldots & 
		 e_2^\rho\left(n\Delta t\right)
	\end{bmatrix}^\prime.
\end{eqnarray}
In the above expressions, $a_0^\omega$, $b_0^\omega$, $\mu_0^\psi$ and $\tau^\psi$ are the parameters of the corresponding prior distributions.  

From the definition~\eqref{eq:psi_omega} we get the following  expression for $\rho$:
\begin{equation}
	\label{eq:rho}
	\rho = \frac{\psi}{\sqrt{\psi^2 + \omega}}
\end{equation}
Thus, having the realisations of $\omega$ and $\psi$ we immediately get the ones for  $\rho$.

\subsubsection{Estimation of $v(t)$}

All the estimation techniques previously presented silently assumed that the values of the volatility process are known. However, the volatility of the asset is a latent variable, i.e. it cannot be directly observed. There are numerous ways of extracting volatility out of the prices of assets, but one of the most effective ones is particle filtering. To make use of it, for each time point $t = k\Delta t$ we need to generate a series of $G$ values called \textit{particles}, denoted by $\{V_j\}_{j=1}^G$. We start by first generating \textit{raw particles} $\{\widetilde{V}_j\}_{j=1}^G$, which will late be resampled to get the final ones. As a first step, we state that $\widetilde{V}_j(0) = \theta_i$ and then we continue generating them for subsequent $k$-s, using 3 auxiliary series:

\begin{equation}
    \varepsilon_j(k\Delta t) \sim \mathcal{N}\left(0,1\right),
\end{equation}
\begin{equation}
	\label{eq:z_eps_pf}
	z_j(k\Delta t)  = \frac{R(k\Delta t) - \mu_i\Delta t -1}{\sqrt{\Delta t}\sqrt{V_j\Big((k-1)\Delta t\Big)}},
\end{equation}
\begin{equation}
	\label{eq:w_eps_pf}
	w_j(k\Delta t)  = z_j(k\Delta t)\rho_i + \varepsilon_j(k\Delta t)\sqrt{1-(\rho_i)^2}.
\end{equation}

Having those three series generated, raw particles can be produced by

\begin{multline}
	\label{eq:V_tilde_pf}
	\widetilde{V}_j(k\Delta t)  = V_j\Big((k-1)\Delta t\Big) + \kappa_i\Bigg(\theta_i - V_j\Big((k-1)\Delta t\Big)\Bigg)\Delta t + \\
	\sigma_i \sqrt{\Delta t}\sqrt{V_j\Big((k-1)\Delta t\Big)} w_j.
\end{multline}

To turn raw particles into proper (final) ones, they need to have weights  $\widetilde{W}$ assigned to them.

\begin{equation}
	\label{eq:W_tilde_pf}
	\widetilde{W}_j(k\Delta t) = 
	    \frac{1}{\sqrt{2\pi\widetilde{V}_j(k\Delta t)\Delta t}} 
	    \exp{\left(-\frac{1}{2} \frac{\Big(R\big((k+1)\Delta t\big) - \mu_i\Delta t -1\Big)^2}{\widetilde{V}_j(k\Delta t) \Delta t}\right)}.
\end{equation}
Those weights then need to be turned to normalized weights $\breve{W}$, so that their sum is equal to 1:

\begin{equation}
	\label{eq:W_bow_pf}
	\breve{W}_j(k\Delta t) = \widetilde{W}_j(k\Delta t)\left(\sum_{j=1}^G{\widetilde{W}_j(k\Delta t)}\right)^{-1}.
\end{equation}

The series of 2-element vectors defined as $\big(\widetilde{V}_j(k\Delta t), \breve{W}_j(k\Delta t)\big)$ can be treated as a categorical distribution. Samples from this distribution are what we consider the \textit{resampled} (final) \textit{particles},

\begin{equation}
	\label{eq:V_ref_pf}
	V_{j}(k\Delta t) \sim \Cat\big(\widetilde{V}_j(k\Delta t), \breve{W}_j(k\Delta t)\big)
\end{equation}

To get the estimate of the volatility, one usually just calculates the average of the resampled particles for each moment of time $k\Delta t$:

 \begin{equation}
	\label{eq:v_pf}
	v(k\Delta t) = \frac{1}{N} \sum_{j=1}^G{V_{j}(k\Delta t)}.
\end{equation}

\subsubsection{Handling of jumps}

The particle filtering mechanism can also be used for estimating jumps in asset prices. For that purpose, two more series need to be generated. The first one $\{J_j\}_{j=1}^G$ will indicate if there was a jump at a given moment of time and the other one $\{Z_j\}_{j=1}^G$ represents the size of the jump. Raw particles can be generated as follows:

\begin{equation}
	\label{eq:J_tilde_pf}
	\widetilde{J}_j(k\Delta t)\sim \mathcal{B}(\lambda^{th}),
\end{equation}
\begin{equation}
	\label{eq:Z_tilde_pf}
	\widetilde{Z}_j(k\Delta t)\sim \mathcal{N}(\mu_0^J, \sigma_0^J),
\end{equation}
where $\lambda^{th}\in[0, 1)$ is an initial proportion of the number of particles which encode the occurrence of a jump to all the particles, $\mu_0^J$ is the assumed mean size of the jump, $\sigma_0^J$ --- of jump's standard deviation. $\mathcal{B}$ denotes a Bernoulli (zero-one) distribution. 

Now, vector $(V_j, J_j, Z_j)$ will be called a single refined particle. Cognately, $(\widetilde{V}_j, \widetilde{J}_j, \widetilde{Z}_j)$ is a raw particle. To turn raw particles into refined ones, weights need to be calculated, but with a slightly different formula this time:

\begin{equation}
	\label{eq:W_tilde_pf_jump}
	\widetilde{W}_j(k\Delta t) = 
	\begin{cases}
	    \begin{aligned}[b]
	    &\frac{1}{\sqrt{2\pi\widetilde{V}_j\big((k-1)\Delta t\big)\Delta t}} \times \\
	    &\exp{\left(-\frac{1}{2} \frac{\big(R(k\Delta t) - \mu_i\Delta t -1\big)^2}{\widetilde{V}_j\big((k-1)\Delta t\big) \Delta t}\right)} 
	    \end{aligned}
	    &\mbox{if } \widetilde{J}_j =0
        \\
        
        \begin{aligned}[b]
        &\frac{1}{\exp{\big(\widetilde{Z}_j(k\Delta t)\big)}\sqrt{2\pi\widetilde{V}_j\big((k-1)\Delta t\big)\Delta t}}
	    \times \\
	    &\exp{\left(-\frac{1}{2} \frac{\Big(R(k\Delta t) - \exp{\big(\widetilde{Z}_j(k\Delta t)\big)}(\mu_i\Delta t +1)\Big)^2}{\exp{\big(2\widetilde{Z}_j(k\Delta t)\big)}\widetilde{V}_j\big((k-1)\Delta t\big) \Delta t}\right)}  
	    \end{aligned}
	    &\mbox{if } \widetilde{J}_j = 1 
    \end{cases}.
\end{equation}

Weights then need to be normalized and the resampling should be done in a way analogous to the case without jumps:

\begin{equation}
	\label{eq:V_pf}
	V_{j}(k\Delta t) \sim \Cat\big(\widetilde{V}_j(k\Delta t), \breve{W}_j(k\Delta t)\big),
\end{equation}

\begin{equation}
	\label{eq:Z_pf}
	Z_{j}(k\Delta t) \sim \Cat\big(\widetilde{Z}_j(k\Delta t), \breve{W}_j(k\Delta t)\big).
\end{equation}
Resampling to get $J_{j}$ can be omitted, as refined values of this series will not be used. Instead, a probability of jump at time $t=k\Delta t$ should be calculated

\begin{equation}
	\label{eq:lambda_k}
	\lambda(k\Delta t)=  \sum_{j=1}^{G} \widetilde{J}_j(k\Delta t)\breve{W}_j(k\Delta t).
\end{equation}
The estimate of parameter $\lambda$ then becomes

\begin{equation}
	\label{eq:lambda_i}
	\lambda_i =  \frac{1}{T}\sum_{k=1}^{n} \lambda (k\Delta t).
\end{equation}

The estimates for the parameters of jump distribution can be calculated as weighted average and weighted standard deviation of sizes of jumps across the time, the weight being the probability of jump:
  
\begin{equation}
	\label{eq:Z_k_pf}
	Z(k\Delta t) = \frac{1}{N} \sum_{j=1}^G{Z_{j}(k\Delta t)}
\end{equation}
\begin{equation}
	\label{eq:mu_J_i}
	\mu_i^J =  \left(\sum_{k=1}^{n} Z(k\Delta t) \lambda(k\Delta t)\right)\left(\sum_{k=1}^{n}{\lambda(k\Delta t)}\right)^{-1}
\end{equation}
\begin{equation}
	\label{eq:sigma_J_i}
	\sigma_i^J =  \sqrt{\left(\sum_{k=1}^{n}{\lambda(k\Delta t)\big(Z(k\Delta t) - \mu_i^J\big)^2}\right)\left(\frac{n-1}{n}\sum_{k=1}^{n}{\lambda(k\Delta t)}\right)^{-1}}.
\end{equation}

Once the jump parameters are estimated, they can be utilised to improve the estimation of the regular parameters, where the effect of the jumps was not taken into account. To do that, one can proceed to the estimation of the singular parameters (or repeat it, if they have already been estimated in a given time step) with the returns redefined as 

\begin{equation}
	\label{eq:Returns_jumps}
	R(k\Delta t) = \frac{S(k\Delta t)}{S\Big((k-1)\Delta t)\Big)} \Bigg(1 - \lambda(k\Delta t) \Big(1-\exp\big(-Z(k\Delta t)\big)\Big)\Bigg).
\end{equation}

\subsubsection{Metaparameters of the framework}

The estimation procedure presented above is dependent on the prior distributions, which need to be assumed in advance. The parameters of those distributions should be picked in a way reflecting our pre-existing beliefs about the values of the parameters we need to estimate. For example, if we know our $\Delta t = 0.01$ and expect the $\mu$ parameter to have a value around $0.5$, by equation \eqref{eq:mu_i} we know that we should use the mean for the prior distribution of $\eta$ to be around $1.005$. The priors which we used are presented in Table \ref{tab:exemplary_priors}.

\begin{table}[h]
\centering
\begin{tabular}{c|c} 
    \textbf{Prior parameter} & \textbf{Value} \\[0.5ex] 
    \hline\hline
    $\mu_0^\eta$ & $1.0004$ \\
    $\sigma_0^\eta$ & $0.001$ \\
    \hline
    $\bm{\Lambda}_0$ & $\begin{bmatrix}
                                10 & 0 \\
                                0 & 5
                        \end{bmatrix}$ \\
    $\bm{\mu}_0$ & $\begin{bmatrix}
                        2\cdot10^{-4} \\
                        0.996
                    \end{bmatrix}$ \\
    \hline        
    $a_0^\sigma$ & $250$ \\
    $b_0^\sigma$ & $0.015$ \\
    \hline        
    $\mu_0^\psi$   & $-0.67$ \\
    $\sigma_0^\psi$ & $0.1$ \\
    $a_0^\omega$ & $1.33$ \\
    $b_0^\omega$ & $0.1$ \\
    \hline        
    $\lambda^{th}$ & $0.15$ \\
    $\mu_0^J$ & $-0.05$ \\
    $\sigma_0^J$ & $0.01$ \\
\end{tabular}
\caption{Priors for the exemplary estimation procedure}
\label{tab:exemplary_priors}
\end{table}

\subsection{Portfolio strategies}
\label{sec:portfolio_strategies}

Part of the experiment described later in this article was related to applying certain portfolio management strategies to assets represented by the trajectories of the Heston Model. We studied two active management strategies, one based on the MACD indicator \cite{appel_technical_2005} and the other one --- on the RSI  indicator \cite{wilder_new_1978}. In essence, both strategies work in the same way. We start by buying a particular number of assets available on the market and then add some free cash to our portfolio. We denote the amount of portfolio cash at time $t$ by $q_0(t)$ and the amounts of the assets by $q_1(t), q_2(t), \ldots, q_N(t)$. Prices of assets are represented by $S_1(t), S_2(t), \ldots, S_N(t)$, with $S_0(t) \equiv 1$ being the value of cash. We then start tracing the values of the trading indicators associated to a given strategy and perform transactions when the indicator sends a "buy" or "sell" signal. For the MACD strategy, trading indicators are of the form 

\begin{equation}
\label{eq:macd_buy_ind}
\mathbbm{1}_{i,p,q,s}^{+}(t) = \begin{cases}
1, &\text{ if } F_{i,p,q,s}(t-\Delta t) < 0 \land F_{i,p,q,s}(t) > 0\\
0, &\text{ otherwise.} 
\end{cases},
\end{equation}

\begin{equation}
\label{eq:macd_sell_ind}
\mathbbm{1}_{i,p,q,s}^{-}(t) = \begin{cases}
1, &\text{ if } F_{i,p,q,s}(t-\Delta t) > 0 \land F_{i,p,q,s}(t) < 0\\
0, &\text{ otherwise.} 
\end{cases},
\end{equation}
where $F_{i,p,q,s}(t) = \EMA_{i, s}(t)-\MACD_{i,p,q}(t)$, in which $\EMA_{i, s}$ is the exponential moving average calculated with the lag $s$ and based on the price process of the $i$-th asset while $\MACD_{i,p,q}$ is an MACD indicator for the price of the $i$-th asset, with short lag $p$ and long lag $q$.

Similar indicators are used in the case of the strategy based on RSI

\begin{equation}
\label{eq:rsi_buy_ind}
\mathbbm{1}_{i,d^+}^{+}(t) = \begin{cases}
1, &\text{ if } \RSI_i(t-\Delta t) < d^+ \land \RSI_i(t) > d^+,\\
0, &\text{ otherwise,} 
\end{cases},
\end{equation}

\begin{equation}
\label{eq:rsi_sell_ind}
\mathbbm{1}_{i,d^-}^{-}(t) = \begin{cases}
1, &\text{ if } \RSI_i(t-\Delta t) > d^- \land \RSI_i(t) < d^-,\\
0, &\text{ otherwise.} 
\end{cases},
\end{equation}
where $\RSI_i$ is the value of the RSI indicator for the $i$-th asset, $d^+$ is the upper boundary level (usually 70) and $d^-$ is the lower boundary level (usually 30).

In case of a "sell" signal (when $\mathbbm{1}^-(t) = 1$) --- a fixed portion $\psi \in [0, 1]$ of the number of assets for which the signal was generated gets sold and the money obtained is turned into portfolio cash. If a "buy" signal appears (when $\mathbbm{1}^+(t) = 1$) --- a fixed portion $\phi \in [0, 1]$ of the amount of the portfolio cash is used to buy assets suggested by the indicator. Those rules are captured by the following set of equations:
\begin{align}
\label{eq:macd_q_update1}
q_i'(t) &= q_i(t-\Delta t) (1-\psi\mathbbm{1}^{-}(t)),\\
\label{eq:macd_q0_update1}
q_0'(t) &= q_0(t-\Delta t) + \sum_{i=1}^{N}{S_i(t)q_i(t-\Delta t)\psi\mathbbm{1}^{-}(t)}.
\end{align}
\begin{equation}
\label{eq:macd_cash_to_buy}
	c(t) = \phi q_0'(t).
\end{equation}
\begin{align}
\label{eq:macd_q_update_final}
q_i(t) &= q_i'(t) +\frac{c(t)}{S_i(t)}\left(\sum_{i=1}^{N}{\mathbbm{1}^{+}(t)}\right)^{-1}\mathbbm{1}^{+}(t),\\
\label{eq:macd_q0_update_final}
q_0(t) &= q_0'(t)-c(t).
\end{align}
where $c(t)$ is the amount of cash exchanged.

\section{Results}
\label{sec:results}

\subsection{Active vs. passive investment debate}
\label{sec:active_passive_debate}

\begin{figure}[h]
     \centering
     \begin{subfigure}[b]{0.3\textwidth}
         \centering
         \includegraphics[width=\textwidth]{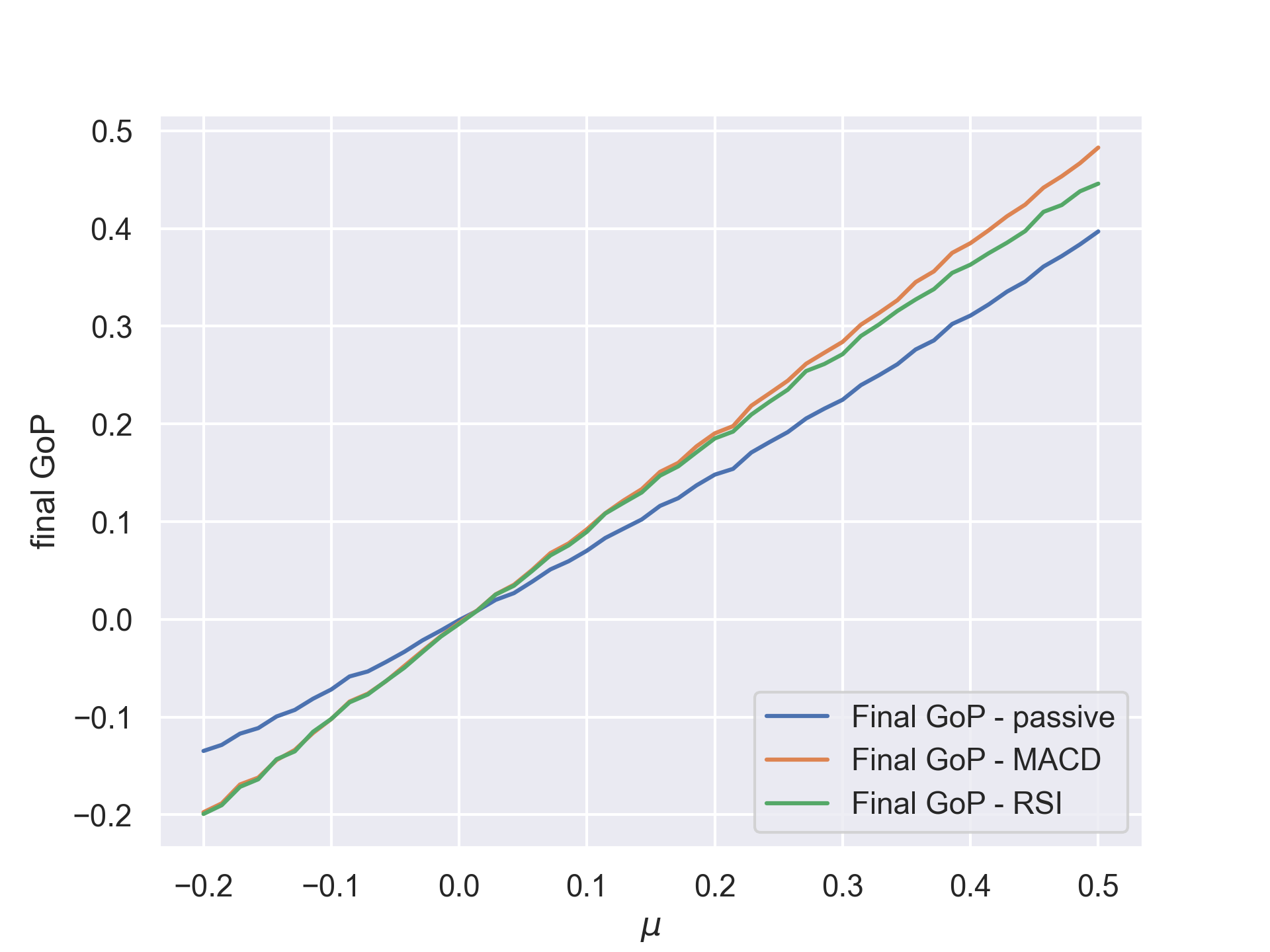}
         \caption{$\lambda = 0$}
         \label{fig:strategy_results_lab=0}
     \end{subfigure}
     \hfill
     \begin{subfigure}[b]{0.3\textwidth}
         \centering
         \includegraphics[width=\textwidth]{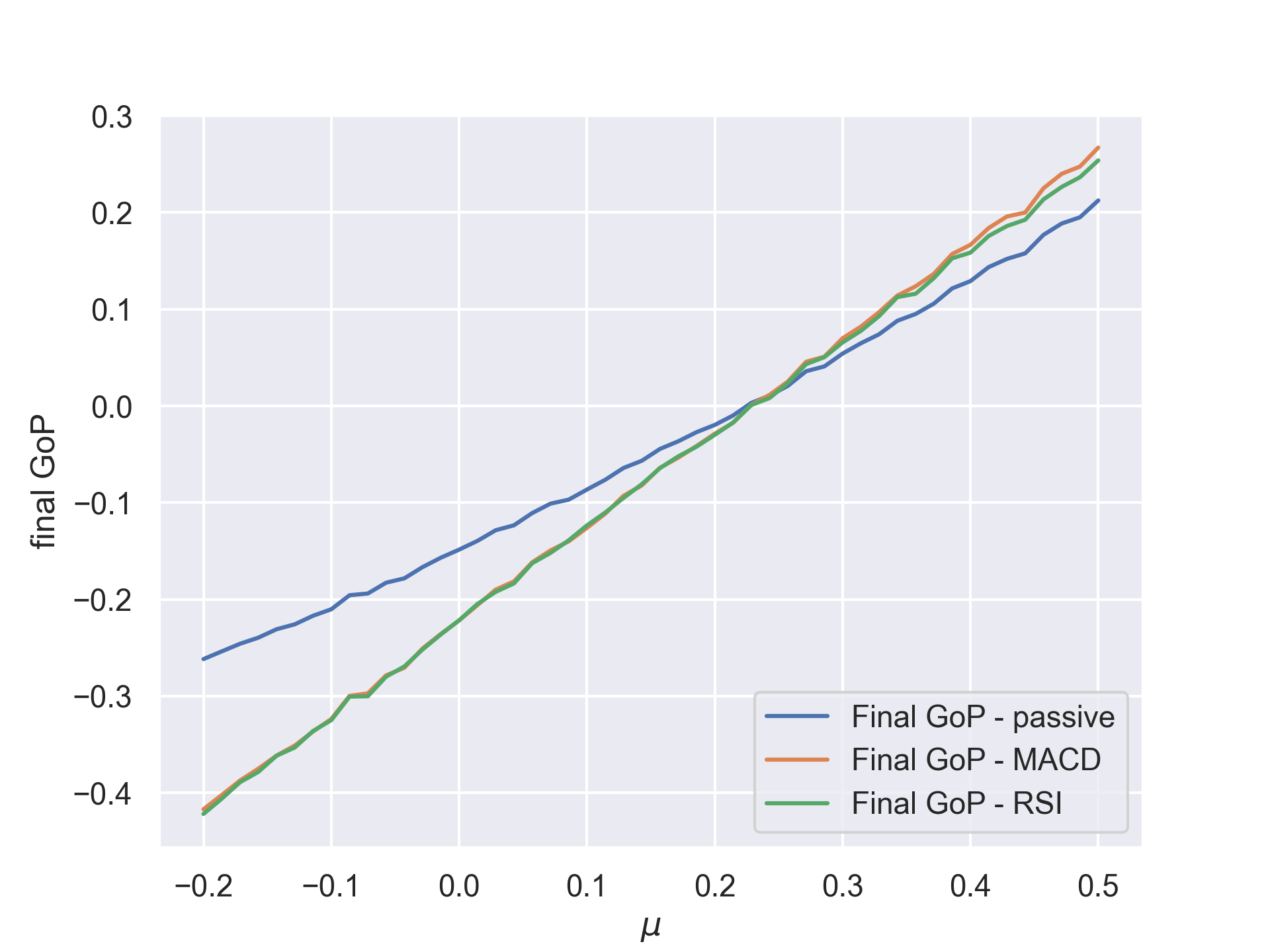}
         \caption{$\lambda = 1$}
         \label{fig:strategy_results_lab=1}
     \end{subfigure}
     \hfill
     \begin{subfigure}[b]{0.3\textwidth}
         \centering
         \includegraphics[width=\textwidth]{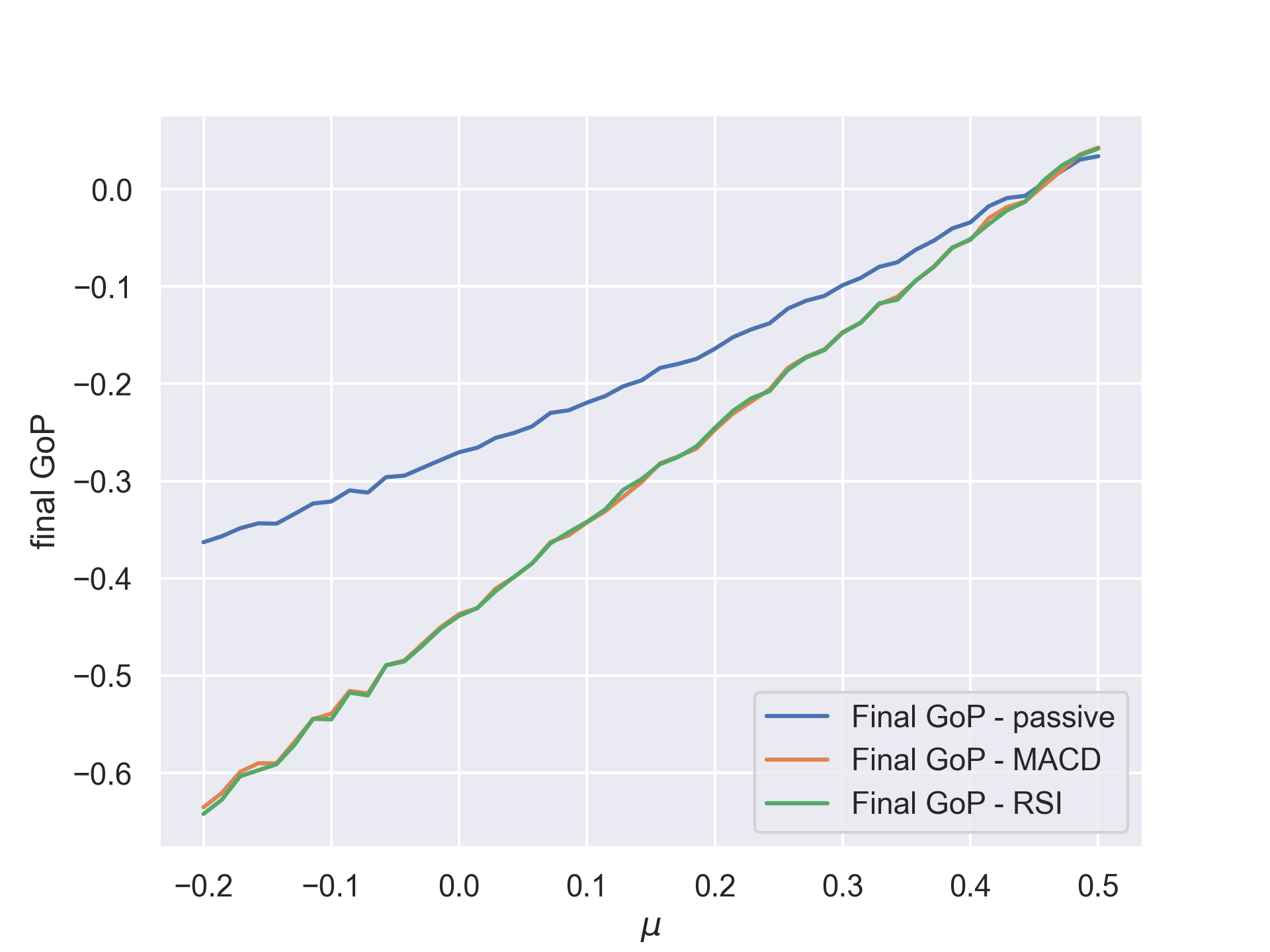}
         \caption{$\lambda = 2$}
         \label{fig:strategy_results_lab=2}
     \end{subfigure}
        \caption{Comparison of the logarithmic growth of portfolio (GoP) led by different investment strategies and with assets of different dynamics. While for the assets with a big tendency to grow and no sudden jumps (big $\mu$, $\lambda=0)$ more aggressive strategies (MACD, RSI) yield better results, the same strategies fall short for assets with lower growth tendency and in presence of jumps (small $\mu$, $\lambda=1,2)$.}
        \label{fig:strategy_results_lambda}
\end{figure}

\begin{figure}[h]
     \centering
     \begin{subfigure}[b]{0.3\textwidth}
         \centering
         \includegraphics[width=\textwidth]{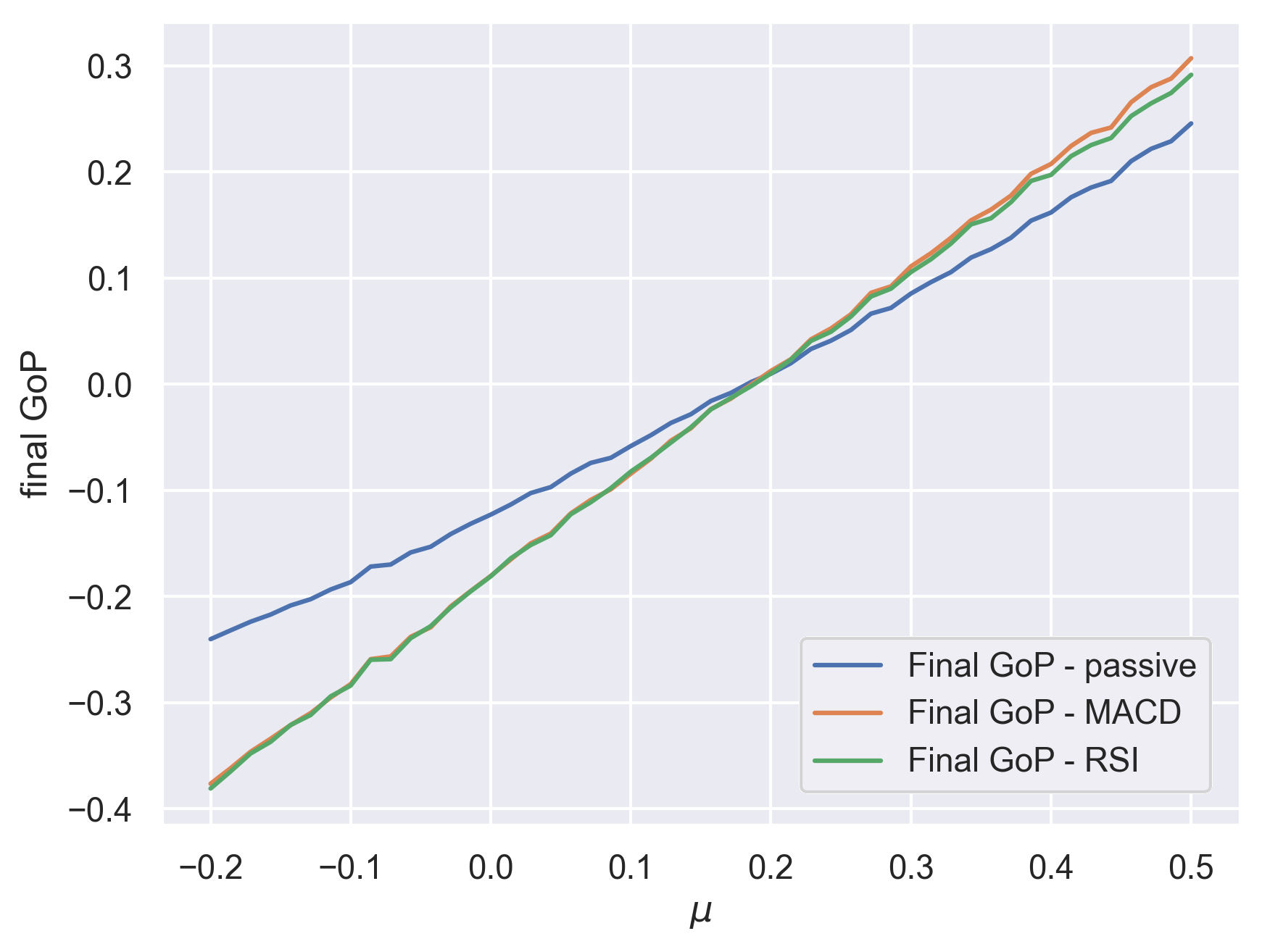}
         \caption{$\mu^J = 0.2$}
         \label{fig:strategy_results_mu_j=02}
     \end{subfigure}
     \hfill
     \begin{subfigure}[b]{0.3\textwidth}
         \centering
         \includegraphics[width=\textwidth]{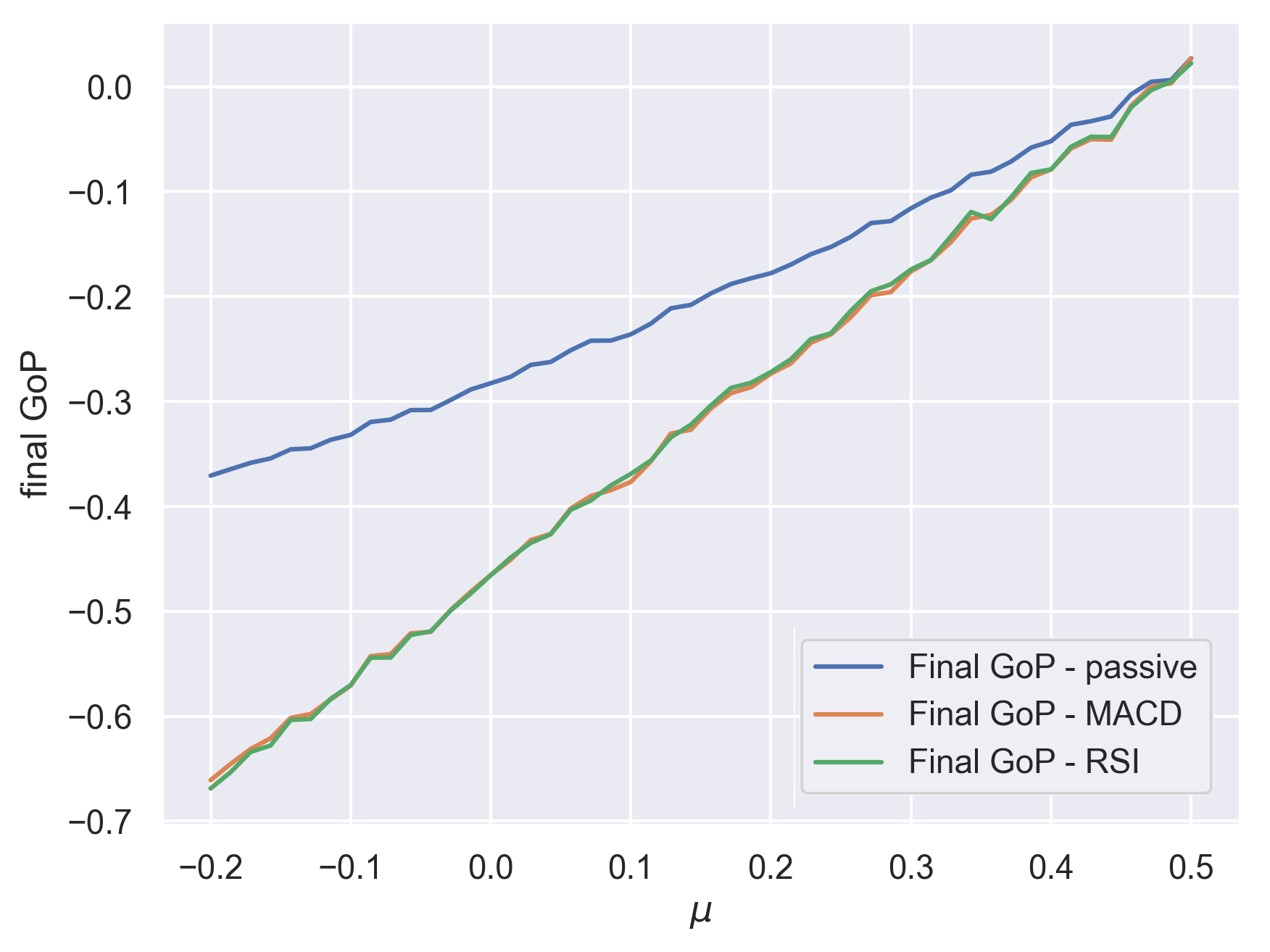}
         \caption{$\mu^J= 0.6$}
         \label{fig:strategy_results_mu_j=06}
     \end{subfigure}
     \hfill
     \begin{subfigure}[b]{0.3\textwidth}
         \centering
         \includegraphics[width=\textwidth]{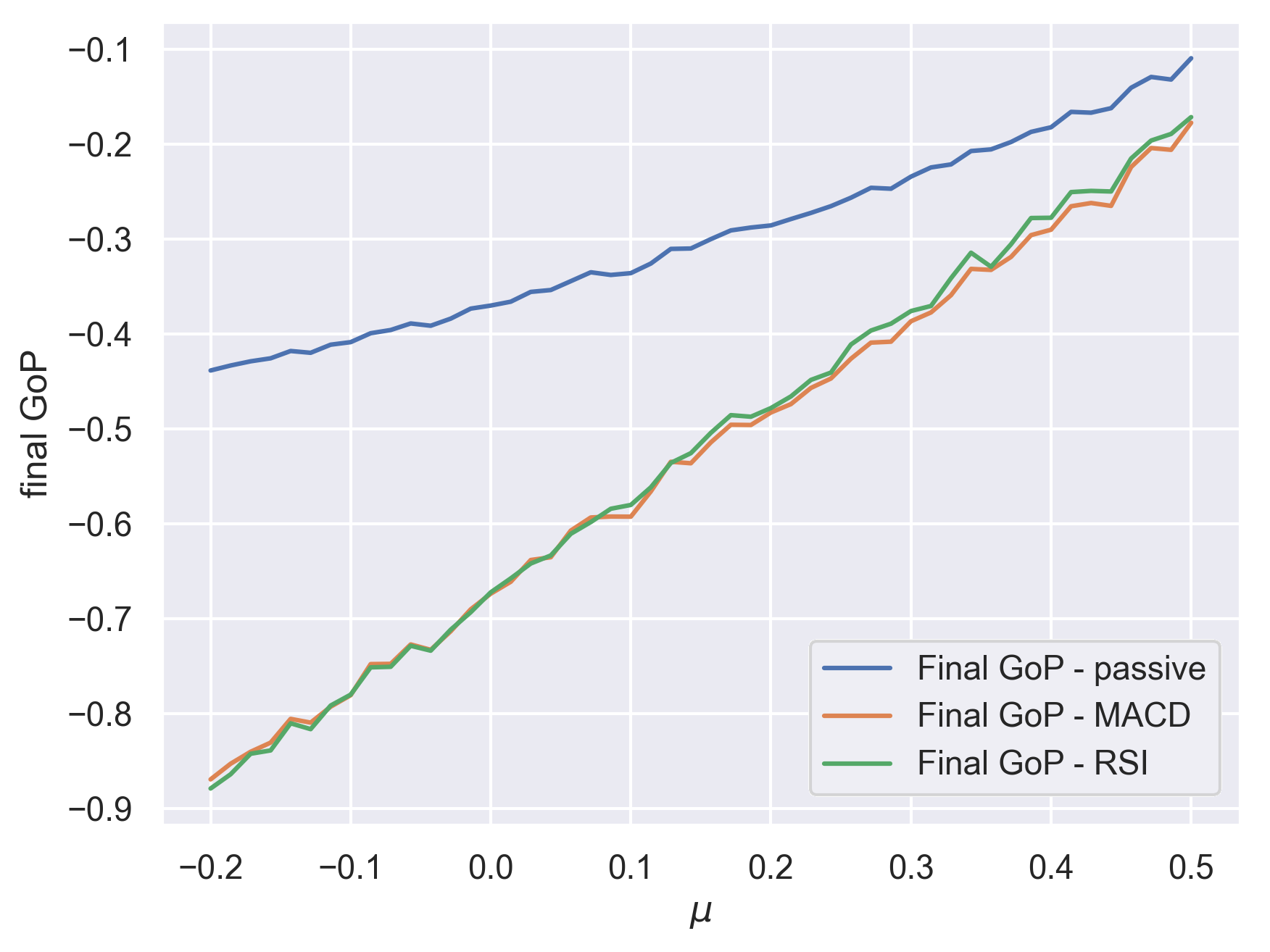}
         \caption{$\mu^J = 1$}
         \label{fig:strategy_results_mu_j=1}
     \end{subfigure}
        \caption{Comparison of the logarithmic growth of portfolio (GoP) led by different investment strategies and with assets of different dynamics. For assets with a big tendency to grow and little jumps (big $\mu$, $\mu^J=0.2)$ more aggressive strategies (MACD, RSI) seem to be applicable in a better way, while the same strategies do not perform well with lower growth tendency and when the jumps are big (small $\mu$, $\mu^J=0.6, 1$).}
        \label{fig:strategy_results_mu_j}
\end{figure}

One of the major findings we present in this paper is that in the debate of whether it is better to invest in an active or passive way, both sides may be right, and one of the key factors determining the success of either of the investing styles are the intrinsic characteristics of the assets which investors trade, namely --- their growth potential. In a series of Monte Carlo experiments, we tested the influence of the parameters of the Heston model on the performance of different investment strategies. Three parameters turned out to be critical for the asset's ability to grow and consequently, for the choice of the strategy: the asset price drift $\mu$, the intensity of jumps $\lambda$, and the size of the jumps $\mu^J$. Generally, the bigger the growth potential (bigger $\mu$, smaller $\lambda$ and $\mu^J$), the more probable it gets that the active portfolio management strategies are going to outperform the passive one. This behaviour has been visualised in Figs.~\ref{fig:strategy_results_lambda} and \ref{fig:strategy_results_mu_j}, for a range of values of $\mu$ and a few values of $\lambda$ and $\mu^J$. The curves represent the value of a portfolio performance measure called \textit{growth of portfolio wealth} (GoP) \cite{alper_effects_2017}, defined as 

\begin{equation}
        \label{eq:gop}
        g(t) = \frac{1}{t} \log\left(\frac{W(t)}{W(0)}\right),
\end{equation}
where $W(t) = \sum_{i=0}^{N}{q_i(t)\cdot S_i(t)}$ is called the portfolio \textit{wealth} (see Ref.~\cite{gruszka_best_2020} for more details). 

For each $\mu$, the value of GoP has been evaluated at $t=T$. Keeping in mind that the assets are generated from the Heston model and hence --- they are random --- for each $\mu$ a number of 1000 portfolios of each kind were generated and the value of $g(T)$, visible on the plot, is an average of all of them. 

To present the dependency between $\mu$, $\lambda$ and portfolio results more comprehensively,  we introduce a new measure ASPI (\textit{Active Strategy Performance Indicator}). It simply assesses whether the difference between the wealth of two investment portfolios, both containing assets characterised by parameters $\mu$ and $\lambda$, is positive. In other words, ASPI is defined as 

\begin{equation}
        \label{eq:ASPI}
        \ASPI(\mu, \lambda) = 
        \left\{
        \begin{array}{ll}
        1 & \text{ if  } W_{\mu,\lambda}^{act}(T) - W_{\mu,\lambda}^{pass}(T) > 0 \\
        0 & \text{ otherwise }
        \end{array}
        \right.\\
\end{equation} 
where $W_{\mu,\lambda}^{act}(T)$ is the terminal wealth (wealth at $t=T$) of an actively managed portfolio, i.e.  

\begin{equation}
        \label{eq:wealth_act}
        W_{\mu,\lambda}^{act} = \sum_{i=0}^{N}{S_i^{\mu, \lambda}(T)\cdot q_i^{act}(T)}.
\end{equation} 
Here, $S_i^{\mu, \lambda}(t)$ is the price of the $i$-th portfolio asset at the moment of time $t$, simulated using drift and jump intensity parameters equal to $\mu$ and $\lambda$, respectively. $q_i^{act}$ is a quantity of this $i$-th asset, as indicated by running the portfolio by an active strategy (like MACD or RSI, as described in section \ref{sec:portfolio_strategies}). Similarly

\begin{equation}
        \label{eq:wealth_pass}
        W_{\mu,\lambda}^{pass} = \sum_{i=0}^{N}{S_i^{\mu, \lambda}(T)\cdot q_i^{pass}(T)},
\end{equation}
where $q_i^{pass}(T) = q_i^{pass}(0) = q_i^{pass}$, as the quantities do not change over time in the passive strategy. 

Also, in this case, the trajectories of the asset prices were random. In such a situation, it is possible, that one pair of portfolios will result in ASPI equal to $1$, and another one, generated for some combination of $\mu$ and $\lambda$ parameters, will give us the ASPI measure of $0$. Hence, to get a full picture, averaging is needed. Therefore, each $(\mu, \lambda)$ combination needs to be used multiple times for the generation of portfolio assets. The average value of ASPI obtained that way can be interpreted as a measure of how much more likely it is for an active strategy to outperform a passive strategy for a given asset type, represented by a $(\mu, \lambda)$ pair. 

\begin{figure}[ht]
\centering
\includegraphics[width=\textwidth]{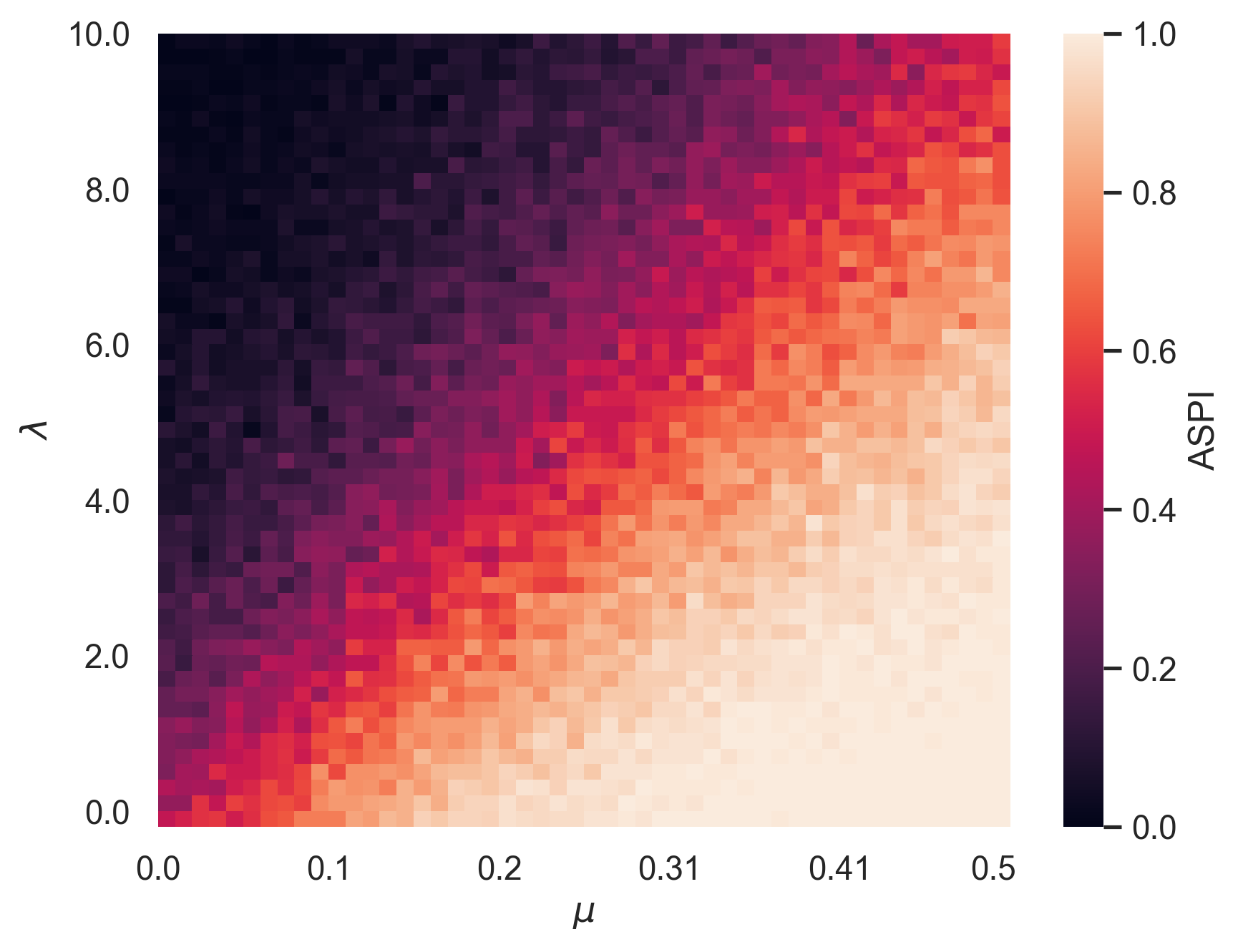}
\caption{A heat map presenting the average value of the ASPI measure Darker colours (values closer to $0$) are an indication that it is better to use a passive strategy, brighter colours (values closer to $1$) --- that the active strategy is preferable.}
\label{fig:ASPI_heatmap}
\end{figure}

We performed such an experiment and presented the results on a heat map in Fig. \ref{fig:ASPI_heatmap}. The x-axis of the map represents the $\mu$ parameter, the y-axis --- parameter $\lambda$ and the color scale is the average ASPI. We used MACD strategy as a model of an actively managed portfolio, since, as Figs.~\ref{fig:strategy_results_lambda} and \ref{fig:strategy_results_mu_j} suggests, both active strategies perform similarly. Darker colours on the heat map, associated with higher frequency of jumps and smaller drift, are an indication that it is better to apply a passive strategy, whereas the bright hues represent the regions where active strategies should be preferable. Identifying proper parameters for an asset and using the map could then allow an investor to select a proper investment strategy based on the character of an asset. In section \ref{sec:real_data_results} we demonstrate that it is indeed possible.

\subsection{Real-data estimation}
\label{sec:real_data_results}

To test our approach, we decided to estimate the parameter values of the Heston model with jumps for the real market data. We selected three well-known market indices for that purpose --- American S\&P500, German DAX and Polish WIG20. We took into consideration the daily closing values of each of the indices, over the time period between the beginning of 2018 and the middle of 2022.

\begin{figure}[ht]
\centering
\includegraphics[width=\textwidth]{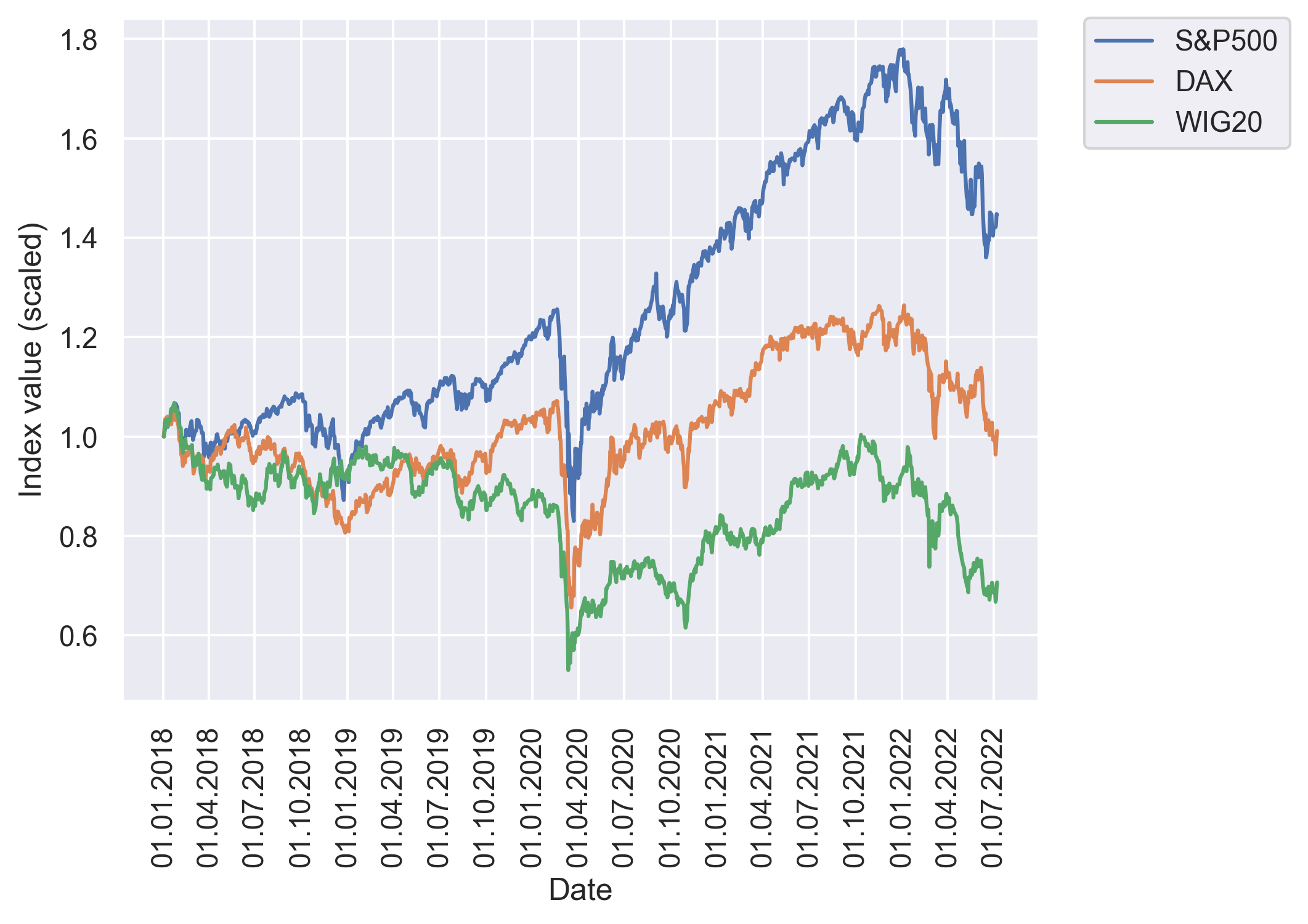}
\caption{A visualisation of the dynamics of three stock market indices --- S\&P500, DAX and WIG20, over the time span of approximately 4.5. The values have been scaled down, so that the initial value of each index is 1, for easier comparison.}
\label{fig:indices_scaled}
\end{figure}

The values all three of the indices have been presented in Fig. \ref{fig:indices_scaled}. It is clearly visible from the plot that the American index has the biggest growth potential. 

In Table \ref{tab:estimation_results}, the results of the parameter estimation for all three of the indices are presented. While the drift parameter $\mu$ turns out to indeed have the greatest value for S\&P500, this index also has the biggest intensity of jumps $\lambda$. Since the average size of a jump is similar for all three indices, the question of what strategy should be used boils down to what are the values of $\mu$ and $\lambda$. Based on the values estimated for S\&P500, this index can be positioned close to the bright area of the heat map in Fig. \ref{fig:ASPI_heatmap}, indicating that active strategies would be more suitable for it. On the other hand, the WIG20 index is located in the darker area of the map which suggest the choice of the passive strategy. It should be noted that the differences in the average ASPI value for the analyzed indices are relatively small. Bigger differences are expected to occur between 
instruments with high tendency to grow combined with very infrequent jumps in prices and the ones with little growth potential and frequent jumps. However, finding such instruments in real financial markets is rather unlikely, as the big growth potential of an instrument is often related to more significant price jumps. But, as we can see from the results of the experiments presented below, even those small differences in the value of $\ASPI$ allow us to aptly choose a proper investment strategy for a given asset.  

\begin{table}
    \centering
    \begin{tabular}{c|c|c|c}
    Parameter & S\&P500 & DAX & WIG20 \\ \hline\hline
    $\mu$ & $0.44$ & $0.30$ & $0.19$\\
    $\kappa$ & $1.17$ & $0.93$ & $0.95$\\
    $\theta$ & $0.06$ & $0.06$ & $0.06$\\
    $\sigma$ & $0.006$ & $0.05$ & $0.006$\\
    $\rho$ & $-0.41$ & $-0.48$ & $-0.40$\\
    $\lambda$ & $8.45$ & $6.45$ & $4.81$\\
    $\mu^J$ & $-0.05$ & $-0.05$ & $-0.05$\\
    $\sigma^J$ & $0.001$ & $0.001$ & $0.001$\\
    \end{tabular}
\caption{Estimation results for selected three stock market indices, under the assumption they follow the Heston model with Merton-style jumps. }
\label{tab:estimation_results}
\end{table}

In our experiment, we applied all three strategies to all three of our indices. The results are presented in Table \ref{tab:strategy_execution_results} and they seem to confirm our findings. The MACD and RSI strategies indeed perform better than the passive portfolio when the asset has bigger growth potential, like in the case of S\&P500. It is also true that for assets performing worse --- like WIG20 --- passive strategies allow if not to earn money, then to lose less than the active ones. 

\begin{table}
    \centering
    \begin{tabular}{c|c|c|c}
    Strategy & S\&P500 & DAX & WIG20 \\ \hline\hline
    passive & $9.48\%$ & $0.63\%$ & $-5.8\%$\\
    MACD-driven & $36.32\%$ & $0.79\%$ & $-12.51\%$\\
    RSI-driven & $32.66\%$ & $3.00\%$ & $-27.30\%$\\
    \end{tabular}
\caption{Returns on investment obtained by executing portfolio management strategies based on investing in one of three selected stock market indices. For the index with the weakest growth potential (WIG20) the passive strategy was the best, allowing to minimise the incurred loss. Investment in the index with the biggest growth potential (S\&P500) earned more when active strategies (utilising MACD and RSI indicators) were utilised.}
\label{tab:strategy_execution_results}
\end{table}

\section{Conclusions}
\label{sec:conclusions}

In the article we presented the relation between certain mathematical characteristics of financial assets and the applicability of particular investment strategies to those assets. By assuming that the prices follow the trajectories of the Heston model with jumps, we concluded that several parameters of this model, mainly the ones related to the growth potential of the stock --- its drift $\mu$ --- and the characteristics of the jumps --- their intensity $\lambda$ and average size $\mu^J$ --- allow us to determine whether following a passive or an active strategy has a bigger profitability potential. It is worth noting that those results were not obtained for a particular, cherry-picked group of instruments. They were the outcomes of a Monte-Carlo-type experiment in which thousands trajectories were included, generated from a well-established mathematical model, proved to accurately reflect the qualities of real-life financial assets. 

Moreover, we estimated the parameters of the Heston model with jumps for three major stock market indices --- S\&P500, DAX and WIG20. The results of the estimation confirmed what was earlier demonstrated via simulations --- that it is possible to select an optimal investment strategy based on the parameters estimated from prices of a given instrument, although it needs to be done with caution, as parameters obtained for most instruments place them near the border between active and passive management strategies on a strategy selection map. This is due to the fact, that a big growth potential (seen as high estimated $\mu$ parameter) is usually paired with high jump intensity (observed as bigger value of $\lambda$). 

Results of our analysis provide a multitude of ways in which the research can be continued. It would be interesting to study if there are any indicators of the applicability of certain investment strategies other than the parameters of the Heston model. Possibly those new indicators could be simpler to obtain than trough the stochastic model estimation process, which is sometimes a challenge both from the perspective of mathematical complexity and computational performance. The range of strategies studied can also be broadened, as the buy-sell investment decisions can be made based on a lot of different factors besides just the indications of the MACD and RSI oscillators, like e.g. a simple measure, especially used by fledgling investors, of when was the last time a given instrument attained a particular price. Making investment decisions is a notoriously challenging topic, so it is of utmost importance to develop proper tools and conduct useful analyses in order to make this process easier and backed by rational arguments at the same time.   

\bibliography{Bibliography}

\begin{thebibliography}{10}

\bibitem{izuddin_impact_2021}
Moch Izuddin.
\newblock The {Impact} {Analysis} {Of} {Fundamental} {Factors} {On} {The}
  {Return} {Of} {Construction} {Company} {Shares}.
\newblock {\em INTERNATIONAL JOURNAL OF ECONOMICS, MANAGEMENT, BUSINESS, AND
  SOCIAL SCIENCE (IJEMBIS)}, 1(1):22--30, January 2021.
\newblock Number: 1.

\bibitem{subrahmanyam_behavioural_2008}
Avanidhar Subrahmanyam.
\newblock Behavioural {Finance}: {A} {Review} and {Synthesis}.
\newblock {\em European Financial Management}, 14(1):12--29, 2008.
\newblock \_eprint:
  https://onlinelibrary.wiley.com/doi/pdf/10.1111/j.1468-036X.2007.00415.x.

\bibitem{hull_options_2018}
John~C. Hull.
\newblock {\em Options, {Futures}, and {Other} {Derivatives}}.
\newblock Pearson, University of Toronto, 10th edition edition, 2018.

\bibitem{stojkoski_geometric_2021}
Viktor Stojkoski, Trifce Sandev, Ljupco Kocarev, and Arnab Pal.
\newblock Geometric {Brownian} motion under stochastic resetting: {A}
  stationary yet nonergodic process.
\newblock {\em Physical Review E}, 104(1):014121, July 2021.
\newblock Publisher: American Physical Society.

\bibitem{stojkoski_ergodicity_2022}
Viktor Stojkoski and Marko Karbevski.
\newblock Ergodicity breaking in wealth dynamics: {The} case of reallocating
  geometric {Brownian} motion.
\newblock {\em Physical Review E}, 105(2):024107, February 2022.
\newblock Publisher: American Physical Society.

\bibitem{vinod_time-averaging_2022}
Deepak Vinod, Andrey~G. Cherstvy, Ralf Metzler, and Igor~M. Sokolov.
\newblock Time-averaging and nonergodicity of reset geometric {Brownian} motion
  with drift.
\newblock {\em Physical Review E}, 106(3):034137, September 2022.
\newblock Publisher: American Physical Society.

\bibitem{haluszczynski_linear_2017}
Alexander Haluszczynski, Ingo Laut, Heike Modest, and Christoph Räth.
\newblock Linear and nonlinear market correlations: {Characterizing} financial
  crises and portfolio optimization.
\newblock {\em Physical Review E}, 96(6):062315, December 2017.
\newblock Publisher: American Physical Society.

\bibitem{valenti_stabilizing_2018}
Davide Valenti, Giorgio Fazio, and Bernardo Spagnolo.
\newblock Stabilizing effect of volatility in financial markets.
\newblock {\em Physical Review E}, 97(6):062307, June 2018.
\newblock Publisher: American Physical Society.

\bibitem{appel_technical_2005}
Gerald Appel.
\newblock {\em Technical {Analysis}: {Power} {Tools} for {Active} {Investors}}.
\newblock Financial Times/Prentice Hall, 2005.

\bibitem{wilder_new_1978}
J.~Welles Wilder.
\newblock {\em New {Concepts} in {Technical} {Trading} {Systems}}.
\newblock Trend Research, 1978.

\bibitem{abbey_is_2012}
Boris~S. Abbey and John~A. Doukas.
\newblock Is {Technical} {Analysis} {Profitable} {forIndividual} {Currency}
  {Traders}?
\newblock {\em The Journal of Portfolio Management}, 39(1):142--150, October
  2012.
\newblock Publisher: Institutional Investor Journals Umbrella.

\bibitem{gruszka_best_2020}
Jarosław Gruszka and Janusz Szwabiński.
\newblock Best portfolio management strategies for synthetic and real assets.
\newblock {\em Physica A: Statistical Mechanics and its Applications},
  539:122938, February 2020.

\bibitem{gruszka_advanced_2021}
Jarosław Gruszka and Janusz Szwabiński.
\newblock Advanced strategies of portfolio management in the {Heston} market
  model.
\newblock {\em Physica A: Statistical Mechanics and its Applications},
  574:125978, July 2021.

\bibitem{lindley_bayes_1972}
D.~V. Lindley and A.~F.~M. Smith.
\newblock Bayes {Estimates} for the {Linear} {Model}.
\newblock {\em Journal of the Royal Statistical Society: Series B
  (Methodological)}, 34(1):1--18, 1972.
\newblock \_eprint:
  https://onlinelibrary.wiley.com/doi/pdf/10.1111/j.2517-6161.1972.tb00885.x.

\bibitem{ohagan_kendalls_1994}
Anthony O'Hagan and Maurice~George Kendall.
\newblock {\em Kendall's {Advanced} {Theory} of {Statistics}: {Bayesian}
  inference. {Volume} {2B}}.
\newblock Edward Arnold, 1994.
\newblock Google-Books-ID: DlrEMgEACAAJ.

\bibitem{gruszka_parameter_2022}
Jarosław Gruszka and Janusz Szwabiński.
\newblock Parameter {Estimation} of the {Heston} {Volatility} {Model} with
  {Jumps} in the {Asset} {Prices}, November 2022.
\newblock arXiv:2211.14814 [q-fin, stat].

\bibitem{heston_closed-form_1993}
Steven Heston.
\newblock A {Closed}-{Form} {Solution} for {Options} with {Stochastic}
  {Volatility} with {Applications} to {Bond} and {Currency} {Options}.
\newblock {\em The Review of Financial Studies}, 6(2):327--343, 1993.

\bibitem{cox_theory_1985}
John~C. Cox, Jonathan~E. Ingersoll, and Stephen~A. Ross.
\newblock A {Theory} of the {Term} {Structure} of {Interest} {Rates}.
\newblock {\em Econometrica}, 53(2):385--407, 1985.
\newblock Publisher: [Wiley, Econometric Society].

\bibitem{merton_option_1976}
Robert~C. Merton.
\newblock Option pricing when underlying stock returns are discontinuous.
\newblock {\em Journal of Financial Economics}, 3(1):125--144, January 1976.

\bibitem{bates_jumps_1996}
David~S. Bates.
\newblock Jumps and {Stochastic} {Volatility}: {Exchange} {Rate} {Processes}
  {Implicit} in {Deutsche} {Mark} {Options}.
\newblock {\em The Review of Financial Studies}, 9(1):69--107, 1996.
\newblock Publisher: [Oxford University Press, Society for Financial Studies].

\bibitem{kloeden_numerical_1992}
Peter Kloeden and Eckhard Platen.
\newblock {\em Numerical {Solution} of {Stochastic} {Differential}
  {Equations}}.
\newblock Stochastic {Modelling} and {Applied} {Probability}. Springer Berlin,
  Heidelberg, Heidelberg, 1 edition, 1992.

\bibitem{johannes_chapter_2010}
Michael Johannes and Nicholas Polson.
\newblock {CHAPTER} 13 - {MCMC} {Methods} for {Continuous}-{Time} {Financial}
  {Econometrics}.
\newblock In YACINE Aït-sahalia and LARS~PETER Hansen, editors, {\em Handbook
  of {Financial} {Econometrics}: {Applications}}, volume~2 of {\em Handbooks in
  {Finance}}, pages 1--72. Elsevier, San Diego, January 2010.

\bibitem{johannes_optimal_2009}
Michael Johannes, Nicholas Polson, and Jonathan Stroud.
\newblock Optimal {Filtering} of {Jump} {Diffusions}: {Extracting} {Latent}
  {States} from {Asset} {Prices}.
\newblock {\em Review of Financial Studies}, 22:2559--2599, June 2009.

\bibitem{christoffersen_volatility_2007}
Peter Christoffersen, Kris Jacobs, and Karim Mimouni.
\newblock Volatility {Dynamics} for the {S}\&{P500}: {Evidence} from {Realized}
  {Volatility}, {Daily} {Returns} and {Option} {Prices}.
\newblock {SSRN} {Scholarly} {Paper} ID 926373, Social Science Research
  Network, Rochester, NY, July 2007.

\bibitem{jacquier_bayesian_2004}
Eric Jacquier, Nicholas~G. Polson, and Peter~E. Rossi.
\newblock Bayesian analysis of stochastic volatility models with fat-tails and
  correlated errors.
\newblock {\em Journal of Econometrics}, 122(1):185--212, September 2004.

\bibitem{alper_effects_2017}
Ofer Alper, Anelia Somekh-Baruch, Oz~Pirvandy, Malka Schaps, and Gur Yaari.
\newblock Effects of correlations and fees in random multiplicative
  environments: {Implications} for portfolio management.
\newblock {\em Physical Review E}, 96(2):022305, August 2017.
\newblock Publisher: American Physical Society.

\end{thebibliography}
\bibliographystyle{unsrt}

\end{document}